\documentclass[final,3p,times,twocolumn]{elsarticle}
\usepackage{amssymb}
\usepackage{amsmath}
\usepackage{tabularx}

\journal{ArXiv}

\usepackage{fancyhdr}
\renewcommand{\headrulewidth}{0pt} % remove header line if desired

\begin{document}

\makeatletter
\def\ps@pprintTitle{%
  \let\@oddhead\@empty
  \let\@evenhead\@empty
  \fancyhf{}%
  \fancyhead[C]{\small \textit{Preprint Version: Submitted version June 2025. \ Copyright: All rights reserved by the Authors.}}%
  \renewcommand{\headrulewidth}{0pt}%
  \pagestyle{fancy}
}
\makeatother

\begin{frontmatter}



\title{Detection of Autonomic Dysreflexia in Individuals With Spinal Cord Injury Using Multimodal Wearable Sensors}


\author[label1,label2]{Bertram Fuchs\fnref{equal}}
\author[label1,label3]{Mehdi Ejtehadi\corref{cor1}\fnref{equal}}
\ead{mehdi.ejtehadi@hest.ethz.ch}
\author[label1,label4]{Ana Cisnal}
\author[label3,label5]{Jürgen Pannek}
\author[label3]{Anke Scheel-Sailer}
\author[label1,label6]{Robert Riener}
\author[label3,label5]{Inge Eriks-Hoogland}
\author[label1,label3]{Diego Paez-Granados\corref{cor1}}
\ead{diego.paez@hest.ethz.ch}
\fntext[equal]{These authors contributed equally as the joint first authors.}

\affiliation[label1]{organization={SCAI-Lab, Department of Health Sciences and Technology (D-HEST), ETH Zurich},%Department and Organization
                    addressline={GLC, Gloriastrasse 37/ 39}, 
                    city={Zurich},
                    postcode={8092}, 
                    country={Switzerland}}
\affiliation[label2]{organization={School of Computation, Information and Technology, Technical University of Munich},%Department and Organization
                    addressline={Arcisstraße 21}, 
                    city={Munich},
                    postcode={80333}, 
                    country={Germany}}
\affiliation[label3]{organization={Swiss Paraplegic Research (SPF)},%Department and Organization
                    addressline={Guido A. Zäch Strasse 4}, 
                    city={Nottwil},
                    postcode={6207}, 
                    country={Switzerland}}
\affiliation[label4]{organization={Institute of Advanced Production Technologies, University of Valladolid},%Department and Organization
                    addressline={Paseo Prado de la Magdalena 3-5}, 
                    city={Valladolid},
                    postcode={470011}, 
                    country={Spain}}
\affiliation[label5]{organization={Swiss Paraplegic Centre (SPZ)},%Department and Organization
                    addressline={Guido A. Zäch Strasse 1}, 
                    city={Nottwil},
                    postcode={6207}, 
                    country={Switzerland}}
\affiliation[label6]{organization={Balgrist University Hospital},%Department and Organization
                    addressline={Forchstrasse 340}, 
                    city={Zurich},
                    postcode={8008}, 
                    country={Switzerland}}
\cortext[cor1]{Corresponding authors.}

% mehdi.ejtehadi@hest.ethz.ch
% bertramfuchs@gmail.com
% cisnal@ieee.org
% juergen.pannek@paraplegie.ch
% anke.scheel-sailer@paraplegie.ch
% robert.riener@hest.ethz.ch
% inge.eriks@paraplegie.ch
% diego.paez@hest.ethz.ch

%% Abstract
\begin{abstract}
%% Text of abstract (maximum  250 words!!)
Autonomic Dysreflexia (AD) is a potentially life-threatening condition characterized by sudden, severe blood pressure (BP) spikes in individuals with spinal cord injury (SCI). 
Early, accurate detection is essential to prevent cardiovascular complications, yet current monitoring methods are either invasive or rely on subjective symptom reporting, limiting applicability in daily file.
This study presents a non-invasive, explainable machine learning framework for detecting AD using multimodal wearable sensors.
Data were collected from 27 individuals with chronic SCI during urodynamic studies, including electrocardiography (ECG), photoplethysmography (PPG), bioimpedance (BioZ), temperature, respiratory rate (RR), and heart rate (HR), across three commercial devices. Objective AD labels were derived from synchronized cuff-based BP measurements. 
Following signal preprocessing and feature extraction, BorutaSHAP was used for robust feature selection, and SHAP values for explainability. We trained modality- and device-specific weak learners and aggregated them using a stacked ensemble meta-model. Cross-validation was stratified by participants to ensure generalizability.
HR- and ECG-derived features were identified as the most informative, particularly those capturing rhythm morphology and variability. The Nearest Centroid ensemble yielded the highest performance (Macro F1 = 0.77±0.03), significantly outperforming baseline models. Among modalities, HR achieved the highest area under the curve (AUC = 0.93), followed by ECG (0.88) and PPG (0.86). RR and temperature features contributed less to overall accuracy, consistent with missing data and low specificity. 
The model proved robust to sensor dropout and aligned well with clinical AD events.
These results represent an important step toward personalized, real-time monitoring for individuals with SCI.
\end{abstract}

% \includegraphics[width=1.0\textwidth{}]{GA_v1.pdf}

% %Research highlights
% \begin{highlights}
% \item First fully non-invasive, BP-ground-truthed, and interpretable ensemble classification system for AD detection in humans using multimodal wearable devices
% \item Results represent an important step toward personalized, real-time AD monitoring to support proactive clinical care and improved quality of life in individuals with SCI
% \end{highlights}

%% Keywords
\begin{keyword}
spinal cord injury \sep autonomic dysregulation \sep autonomic dysreflexia \sep artificial intelligence \sep multimodal data fusion \sep wearable sensors \sep mobile health


\end{keyword}

\end{frontmatter}



\section{Introduction}
\label{intro}

\subsection{Background}
Spinal cord injury (SCI) is a life-altering condition affecting approximately 52.5 individuals per million globally each year~\cite{priceThermoregulationFollowingSpinal2018}, with over 58\% of cases resulting in cervical-level injuries~\cite{singhGlobalPrevalenceIncidence2014}. Beyond motor and sensory impairments, SCI frequently disrupts autonomic cardiovascular control, giving rise to autonomic dysregulation (ADys), a term encompassing both autonomic dysreflexia (AD) and orthostatic hypotension (OH)~\cite{krassioukovAssessmentAutonomicDysfunction2007}.

AD affects up to 90\% of individuals with injuries above the T6 level~\cite{allenAutonomicDysreflexia2022}, characterized by rapid, often severe increases in systolic blood pressure (SBP) that can result in strokes, seizures, or even death if not treated~\cite{sureshAutomatedDetectionSymptomatic2020}. In contrast, OH involves a significant drop in blood pressure (BP) upon standing, leading to fatigue, dizziness, and reduced quality of life~\cite{bradleyOrthostaticHypotension2003}. Both conditions contribute to elevated cardiovascular risk in SCI individuals, especially during the early months post-injury when episodes frequently begin to manifest~\cite{lindanIncidenceClinicalFeatures1980}. Prompt detection and intervention are therefore essential.

Despite the clinical importance, ADys episodes often go undetected outside of hospital environments due to subtle or absent perceivable symptoms and limited awareness among caregivers~\cite{allenAutonomicDysreflexia2022}. Current diagnosis is primarily based on invasive monitoring or subjective patient-reported symptoms, which limits scalability and reliability in everyday settings. An objective, automated, and non-invasive detection system based on wearable sensors could enable earlier identification and timely response to ADys episodes, improving outcomes for the SCI population.

\subsection{Prior Work}
AD remains a critical and under-monitored complication for individuals with SCI, especially those with lesions above T6. Existing studies have explored both human and animal models to develop automated AD detection systems using machine learning.

Suresh et al. \cite{sureshAutomatedDetectionSymptomatic2020} proposed an early wearable-based system leveraging electrodermal activity (EDA), heart rate (HR), and skin temperature (Temp) sensors from a wrist-worn smartwatch. Their machine learning model (SVM) trained on self-reported AD events by SCI participants achieved high accuracy, but relied on a small sample of seven participants and lacked systematic cross-validation. In a follow-up study \cite{sureshAutomaticDetectionCharacterization2022}, the same group extended their multimodal sensing approach, incorporating a more complex telemetry system and increasing the participant count to eleven. They reported a detection accuracy of 94.1\%, emphasizing real-time application feasibility in community settings. Although lack of cross validation methods and small sample size, limits the generalizability of their findings.

Sagastibeltza et al. \cite{sagastibeltzaPreliminaryStudyDetection2022} investigated AD onset through controlled bladder filling in a clinical environment with only five human subjects. Their approach uniquely combined physiological measurements (e.g., ECG, peripheral resistance) with hormone analysis and patient history, but it remained a preliminary proof-of-concept without wearable deployment.

Pancholi et al. \cite{Pancholi_2024} proposed a novel approach using deep neural networks (DNNs) trained on skin nerve activity (SKNA) data collected from a rat model via colorectal distension to induce AD. Although their system achieved high accuracy (93.9\% $\pm$ 2.5\%) and low false-negative rates, the use of invasive SKNA signals and non-human subjects limits the clinical translatability.

Lastly, Suresh et al. \cite{sureshFeatureSelectionTechniques2022} conducted an in-depth analysis of feature selection techniques to optimize AD detection from physiological signals. This study focused on algorithmic refinement rather than system-level deployment, further underscoring the fragmented and exploratory state of current AD detection research.

These prior works highlight the potential of multimodal biosignal analysis, particularly using EDA and Temp; however, they also expose critical limitations: reliance on subjective ground truths, lack of temporal detection, and limited generalizability to everyday environments, with EDA and skin temperature being highly prone to environmental error. 
Currently, no study or method has demonstrated automated detection of AD episodes using non-invasive wearable sensor data, and validated it with objective BP measurements.

\subsection{Goal of This Study}
In this study, we addressed this gap by investigating what wearable biosignals carry the most predictiveness for detecting AD. We hypothesize that these events are associated with unique physiological changes in each person, including alterations in Temp, EDA, heart rate variability (HRV), and respiratory rate (RR) ~\cite{sureshAutomatedDetectionSymptomatic2020, bradleyOrthostaticHypotension2003}. Therefore, a highly non-linear relationship across signals is expected and exploited using ensemble machine learning classifiers.

The contributions of this work are three-fold:  
(1) Identification of informative signal modalities and physiological features descriptive of AD;  
(2) Development of the first ensemble classifier for wearable, multi-modal AD detection based on objectively measured BP variations; 
(3) Design of a robust ensemble classification framework with BorutaSHAP-based feature selection, capable of maintaining performance under sensor noise or modality failure, ensuring practical applicability in real-world settings.

\section{Methods}
\label{sec:methods}

\subsection{Study Population}

This observational proof-of-concept study included 27 individuals with chronic, motor- and sensory-complete SCI (AIS A) with level of injury (LOI) at or above the T6 level. Participants were recruited from inpatient and outpatient services at the Swiss Paraplegic Center (SPC), where they underwent a scheduled urodynamic study (UDS). Data were compiled from three ethically approved research protocols.

The compiled dataset was screened for completeness, and subjects with insufficient ECG, PPG, or reference BP measurements were excluded from analysis. Demographic characteristics and sensor data availability after screening are presented in Section~\ref{sec:results}.

\subsection{Clinical Protocol and Reference AD Labels}
During UDS, the bladder was filled with 37\textdegree C saline solution to induce AD. SBP, DBP, and HR were recorded every 2--3 minutes using a medical-grade BP cuff. 
Reference AD episodes were annotated using clinical guidelines from Krassioukov et al.~\cite{krassioukovAssessmentAutonomicDysfunction2007} and the American Autonomic Society~\cite{ConsensusStatementDefinition1996}, which define AD as a sustained increase in SBP of $\geq$20 mmHg above baseline. The baseline was computed as the average of the first three SBP measurements obtained during the UDS.  A piecewise-cubic Hermite interpolating polynomial (PCHIP) was fitted to the sparse BP data to generate continuous reference labels.
Figure~\ref{fig:sensory_setup} illustrates a representative AD episode with interpolated SBP, AD threshold, and bladder filling.

\subsection{Wearable Sensor System}
\begin{figure*}[tb]
    \centering
    \includegraphics[width=1.0\linewidth]{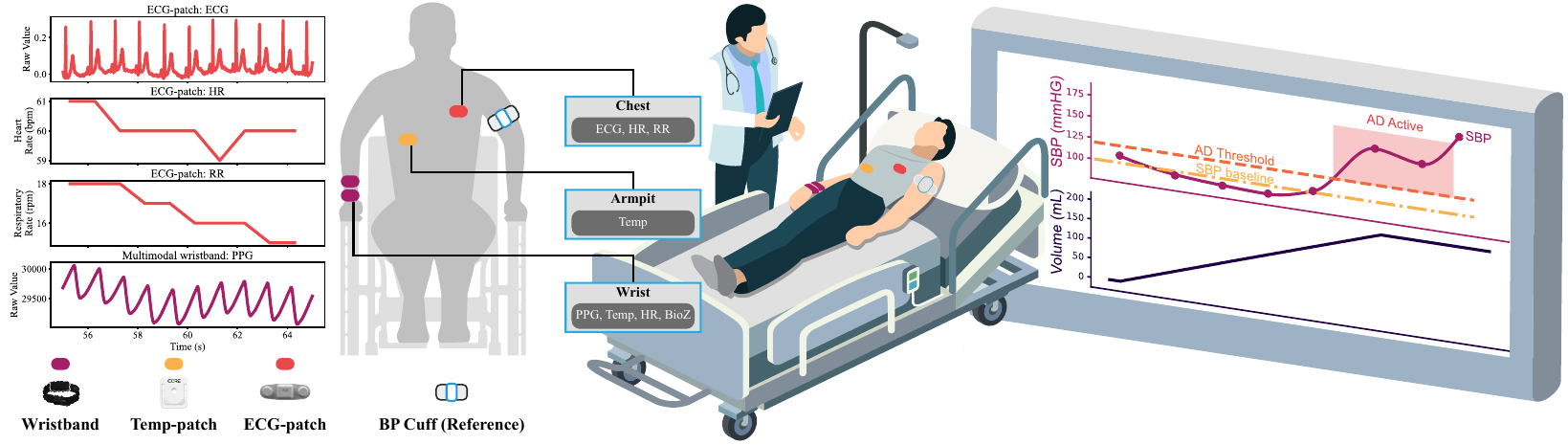}
    \caption{Sensor placement and modality mapping. Two multimodal wristbands captured PPG, skin Temp, RR (front), and BioZ (back); a chest-mounted ECG-patch recorded ECG, HR, and RR; A temperature patch measured torso Temp; and a medical cuff provided reference BP. The right panel illustrates UDS annotation with example bladder volume and SBP plots, where AD onset is marked by an SBP rise $\geq$20\,mmHg above baseline. The left panel shows representative raw signals from one subject: ECG, HR, and RR from ECG-patch, and PPG from multimodal wristband.}
    \label{fig:sensory_setup}
\end{figure*}

A multimodal wearable system recorded photoplethysmography (PPG), electrocardiography (ECG), bio-impedance (BioZ), skin and core body temperature (Temp), HR, and respiratory rate (RR). Two multimodal wristbands (CardioWatch, Corsano) captured PPG, skin temperature, RR, and BioZ. A chest-mounted ECG-patch (Wearable ECG Monitor, VivaLNK) recorded ECG, HR, and RR. A Temp-patch (CORE, greenTEG) provided both skin and core body temperature. Table~\ref{tab:modalities} and Figure~\ref{fig:sensory_setup} detail device placement and sampling specifications. The figure also illustrates sample signal traces from selected device modalities.

\begin{table}[t]
\centering
\caption{Sensory setup and device specifications used for the classification of AD.}
\begin{tabular}{p{0.20\linewidth} p{0.35\linewidth} p{0.30\linewidth}}
\hline
Device ID & Device Model & Modalities \\
\hline
Multimodal wristband &
CardioWatch 287-2B (Corsano Health Inc., Cambridge, MA, USA) &
PPG (128 Hz), Temp (1/60 Hz), RR (1 Hz), BioZ (32 Hz) \\

ECG-patch &
Wearable ECG Monitor
(VivaLNK, Campbell, CA, USA) &
ECG (128 Hz), HR (1 Hz), RR (1 Hz) \\

Temp-patch &
CORE (greenTEG, Zurich, CH) &
Temp (1/60 Hz) \\ \hline
\end{tabular}
\label{tab:modalities}
\end{table}

\subsection{Preprocessing and Feature Extraction}
Each biosignal underwent modality-specific preprocessing and feature extraction using sliding windows with (duration, step) configurations: (5\,s, 5\,s), (10\,s, 5\,s), (10\,s, 10\,s), (30\,s, 10\,s), (60\,s, 5\,s), and (60\,s, 10\,s).
Signals were synchronized across devices, denoised, and quality-controlled as needed. Features were computed per window to capture 
patterns relevant to AD detection. For ECG and PPG, additional template-based features were 
extracted from beat-to-beat waveforms, resulting in both window-level and beat-level 
descriptors. Table~\ref{tab:feat_summary} summarizes preprocessing and feature types per 
modality. The final set includes time-domain, statistical, and frequency-domain features, 
derived using custom methods and open-source libraries (VitalPy~\cite{cisnal_2024}, 
NeuroKit2~\cite{Makowski_2021}, BIOBSS~\cite{BIOBSS}).

\begin{table*}[h!]
\caption{Summary of Preprocessing and Feature Extraction per Modality. Abbreviations: AI -- Augmentation Index; CT -- Crest Time; DW -- Diastolic Width; FFT -- Fast Fourier Transform; HRV -- Heart Rate Variability; LASI -- Large Artery Stiffness Index; LBP -- Local Binary Patterns; LF/HF -- Low-/High-Frequency ratio; MAD -- Median Absolute Deviation; NPV -- Normalized Pulse Volume; pNN50 -- Proportion of NN intervals \(>50\,\text{ms}\); RI -- Reflection Index; RMSSD -- Root Mean Square of Successive Differences; RMS -- Root Mean Square; SCL -- Skin Conductance Level; SCR -- Skin Conductance Response; SDNN -- Standard Deviation of NN intervals; SNR -- Signal-to-Noise Ratio; SW -- Systolic Width.}
\label{tab:feat_summary}
\centering
\begin{tabularx}{\linewidth}{>{\hsize=0.08\hsize}X>{\hsize=0.45\hsize}X>{\hsize=0.47\hsize}X}
\hline
Modality & Preprocessing & Features Extracted \\
\hline
PPG & Signal inversion (sensor reflection correction), 4th-order Butterworth bandpass (0.25--10\,Hz), baseline correction via iteratively reweighted least squares~\cite{zhang_baseline_2010}, heartbeat segmentation with derivative-based adaptive thresholding~\cite{cisnal_2024}, systolic peak normalization, and SQI filtering (skewness, kurtosis, SNR) with subject-specific thresholds & Template-based features: fiducial timing, amplitudes, slopes, areas, and indices (RI, AI, CT, NPV, LASI); statistical: mean, variance, MAD, RMS, skewness, kurtosis, entropy, perfusion; frequency: FFT harmonics, spectral stats (e.g., skewness, energy, wavelet entropy); additional temporal/spectral descriptors from VitalPy~\cite{cisnal_2024} and BIOBSS~\cite{BIOBSS}, including DW, SW, DW/SW ratios, amplitude percentiles, and peak FFT amplitudes \\

ECG & Wavelet denoising (bior4.4), sequential median filtering (200/600\,ms) to correct P/QRS/T baseline drift~\cite{Zhancheng_2014}; R-peak detection using Christov's method~\cite{Christov_2004}; outlier beats removed via morphology and amplitude thresholds from baseline & Template-based: fiducial durations (QRSw, QSd), amplitude differences (PQa, QRa, RSa), amplitude ratios~\cite{SaenzCogollo_2020}; morphology descriptors: wavelets (db3), Hermite coefficients, LBP patterns~\cite{MondejarGuerra_2019}; time-series: HRV (mean HR, SDNN, RMSSD, pNN50), frequency (LF/HF via Welch), Poincar\'{e} (SD1, SD2)~\cite{Makowski_2021} \\

BioZ & Median filter (4\,s) to separate tonic (SCL) and phasic (SCR) components~\cite{Makowski_2021} & Separate features for tonic (SCL) and phasic (SCR) components: statistical (mean, std, min/max, derivatives), SCR descriptors (peak count, amplitude, rise time, area)~\cite{Zhang_2018}; bandpower (energy, variance, power in 0--0.5\,Hz bands); spectral (magnitude area, freq. mean/std, range, skewness, kurtosis) \\

Temp & None (low sampling rate) & Mean absolute value; 1st--3rd derivatives over 1-, 2-, and 3-minute intervals to capture temporal trends~\cite{Choi_2012}. Missing features are replaced by propagating the last valid feature. \\

HR & Normalized to resting; missing values linearly interpolated & Time-domain features (e.g. mean RR, SDNN, RMSSD, pNN50); frequency-domain features using cubic interpolation at 4~Hz (e.g. LF, HF, LF/HF, total power); nonlinear features (Poincar\'{e} SD1, SD2) \\

RR & Normalized to resting & Same as Temp \\
\hline
\end{tabularx}
\end{table*}

\subsection{Machine Learning Framework}

\subsubsection{SHAP-Based Feature Importance and Selection}
Feature importance and selection were addressed jointly using TreeSHAP~\cite{Lundberg_2017} and BorutaSHAP~\cite{borutashap}. TreeSHAP computes exact per-sample Shapley values \(\phi_i^{(j)}\) from XGBoost ensembles, representing the marginal contribution of feature \(i\) to the prediction \(f(x^{(j)})\) for sample \(j\), relative to the expected model output \(\mathbb{E}[f(x)]\). Global rankings used mean absolute SHAP values, z-scored across features; local relevance used standardized distributions of \(\phi_i^{(j)}\).

For selection, raw features were first z-normalized. Then SHAP values were standardized:

\begin{equation*}
\bar\phi_i = \frac{1}{N} \sum_{j=1}^N \phi_i^{(j)}
\end{equation*}

\begin{equation*}
s_i = \sqrt{ \frac{1}{N-1} \sum_{j=1}^N \left(\phi_i^{(j)} - \bar\phi_i\right)^2 }
\end{equation*}

\begin{equation*}
z_i^{(j)} = \frac{\phi_i^{(j)} - \bar\phi_i}{s_i}
\end{equation*}

BorutaSHAP compared these standardized SHAP scores against  “shadow” (permuted) features, retaining only features that performed consistently better than the shadow features. Selection was conducted both locally (per modality/device) and globally. XGBoost with column subsampling mitigated multicollinearity. Tentative features were treated as rejected after 500 iterations. To reduce label noise, only samples within 2 minutes of a reference BP measurement were considered.

\subsubsection{Ensemble Architecture and Evaluation}
To ensure robust AD detection under real-world conditions (e.g., sensor dropout or signal degradation), we implemented a multi-modal ensemble framework. The architecture comprises multiple Random Forest weak learners, each trained on features from one modality (e.g., ECG, PPG, BioZ) or device (e.g., ECG-patch, multimodal wristband, Temp-patch). Inputs were standardized via a RobustScaler and limited to features selected by BorutaSHAP. Each learner was trained on class-balanced data via random undersampling of the majority class.\\

Two aggregation strategies were tested: (1) $k$-threshold voting, which predicts AD if at least \(k\) weak learners concur; and (2) stacked ensemble learning, where outputs from weak learners were used to train over 40 classifiers (e.g., Random Forest, XGBoost, Logistic Regression).

\begin{figure}[tb]
    \centering
    \includegraphics[width=\linewidth]{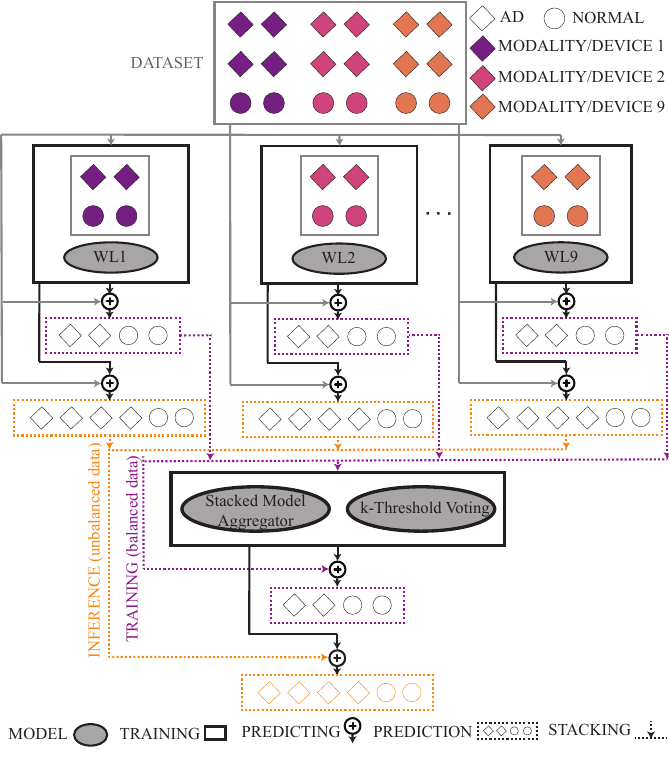}
    \caption{Ensemble classifier. Each of the 9 weak learners (WLi) is trained on features from a single one of the 6 modalities or 3 devices. Final predictions are aggregated via (1) $k$-threshold voting or (2) a stacked ensemble meta-learner. Training uses balanced data; validation is carried out on imbalanced sets.}
    \label{fig:ensemble_pipeline}
\end{figure}

Model generalization was assessed using Leave-One-Subject-Out cross-validation. Evaluation on held-out subjects used the original class imbalance to reflect clinical prevalence. To address sampling variability, training and testing were repeated 10 times with different random seeds. Figure~\ref{fig:ensemble_pipeline} demonstrates the ensemble classifier.

% \subsection{Evaluation Metrics}
Performance was evaluated via ROC-AUC and macro F1-score, accounting for class imbalance. Class-wise precision, recall, and F1 were also calculated and macro averaged.

\section{Results}
\label{sec:results}

\subsubsection{Data}

\begin{figure}[tb]
    \centering
    \includegraphics[width=1.0\linewidth]{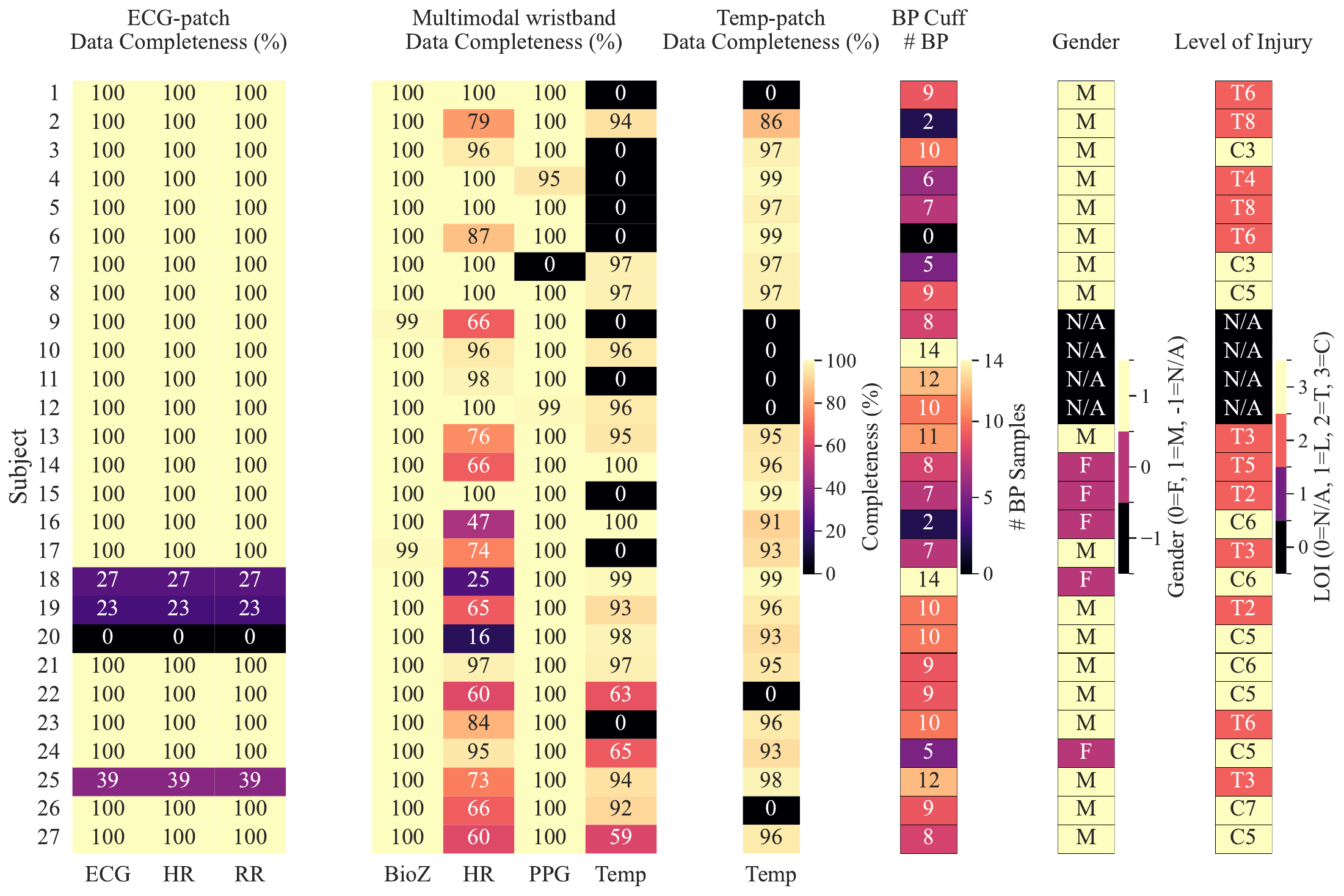}
    \caption{Participant overview and data completeness. The heatmap shows gender, LOI, number of reference BP samples, and sensor data completeness (\%) for participants, across modalities and devices: ECG, HR, RR (ECG-patch); BioZ, HR, PPG, Temp (multimodal wristband); Temp (Temp-patch). Subjects with missing ECG/PPG or insufficient BP samples were excluded from AD classification. C – Cervical; T – Thoracic; L – Lumbar.}
    \label{fig:data}
\end{figure}

As shown in Figure~\ref{fig:data}, 27 participants were enrolled. Gender and level of injury (LOI) data were available for 23 individuals (18 male, 5 female; 12 thoracic, 11 cervical). One thoracic case (T8) was retained despite being below the T6 threshold due to clinically confirmed AD episodes.

Ten participants were excluded from AD classification: five due to incomplete ECG or PPG data (subjects 7, 18–20, 25), three due to insufficient reference BP measurements (subjects 2, 6, 16), and two (subjects 22, 26) due to motion-induced ECG/PPG signal degradation. This resulted in 17 participants being used for model training and evaluation.

These 17 participants had an average of $8.58 \pm 3.01$ BP references recorded over $21.09 \pm 6.27$ minutes of UDS. Among them,  7 exhibited AD episodes, consistent with previous findings~\cite{Krassioukov_2023} reporting AD incidence rates of 37–78\% during UDS, depending on neurological level and SCI severity.

Using the selected $(60\,\mathrm{s},\,10\,\mathrm{s})$ sliding window configuration, the final dataset included 3,696 samples (210 AD, 3,486 normal), averaging $246.4 \pm 64.5$ sam

\subsection{Sliding Window Size Analysis}
\begin{figure}[tb]
\centering
\includegraphics[width=\linewidth]{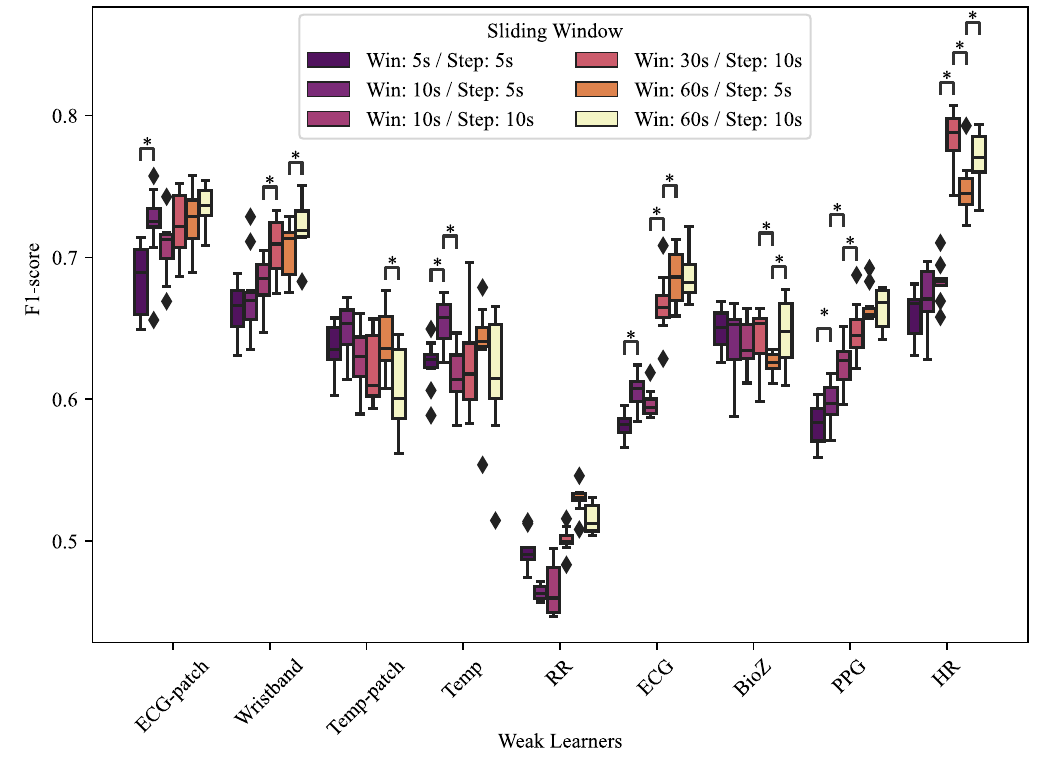}%
\caption{F1-score of individual weak learners (by modality and device) across different sliding window configurations. Asterisks (*) indicate statistically significant differences between consecutive window configurations (paired t-test, $p<0.05$).}
\label{fig:windows_analysis}
\end{figure}

Figure~\ref{fig:windows_analysis} shows the F1-score per weak learner and configuration for the 6 evaluated windowing settings. To assess overall performance, we also averaged F1-scores across all weak learners: $0.62 \pm 0.12$ (5\,s, 5\,s), $0.65 \pm 0.12$ (10\,s, 5\,s), $0.62 \pm 0.12$ (10\,s, 10\,s), $0.67 \pm 0.12$ (30\,s, 10\,s), $0.67 \pm 0.12$ (60\,s, 5\,s), and $0.68 \pm 0.12$ (60\,s, 10\,s).
Based on this analysis, (60\,s, 10\,s) was selected as the optimal configuration and is used in the subsequent analyses.

\subsection{Feature Importance and Feature Selection}

\begin{figure}[tb]
\centering
\includegraphics[width=\linewidth]{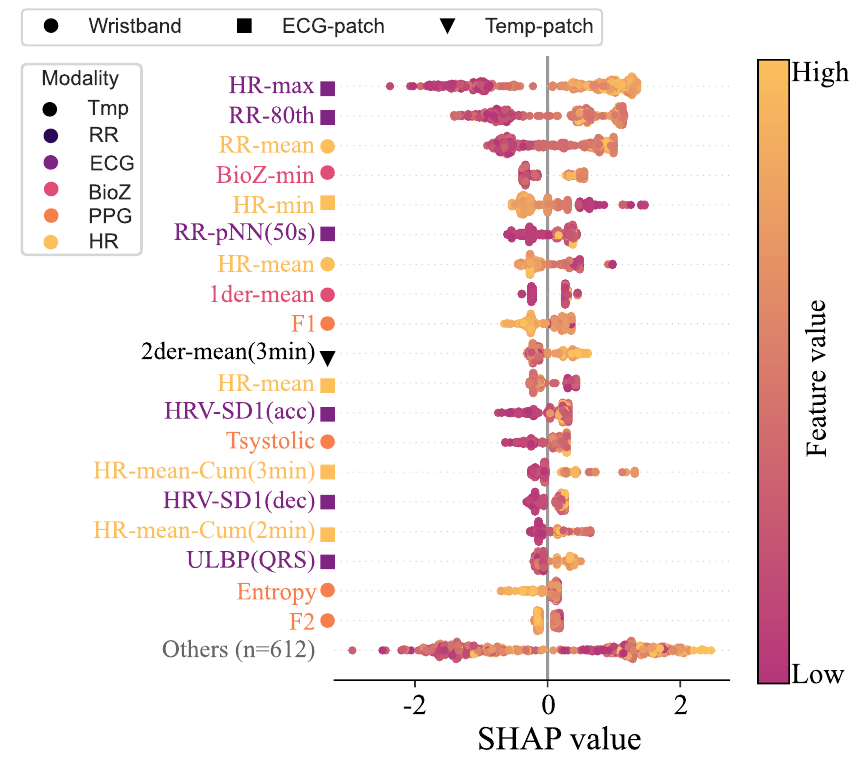}
\label{fig:shap-unbalanced}
\caption{SHAP summary plot of the top features for AD detection using an XGBoost classifier trained on the original imbalanced dataset. Features are ranked by their mean absolute Shapley values. Text colors indicate signal modality; markers indicate device.}
\label{fig:shap-analysis}
\end{figure}

Figure~\ref{fig:shap-analysis} ranks features by Shapley importance, aiding physiological interpretation. ECG- and HR-derived features dominated, including HR stats (mean, min), cumulative mean HR (\textit{HR-mean-Cum}), and HRV metrics (\textit{RR-pNN(50s)}, \textit{RR-80th}, \textit{RR\_maxNN}). Morphological ECG features like \textit{ULBP(QRS)} (a QRS complexity descriptor based on Uniform Local Binary Patterns) also appeared. BioZ contributed with SCL \textit{max} and \textit{mean}; PPG yielded \textit{F1} (fundamental spectral frequency). No Temp and RR-derived features ranked among the top.

\begin{figure}[tb]
    \centering
    \includegraphics[width=\linewidth]{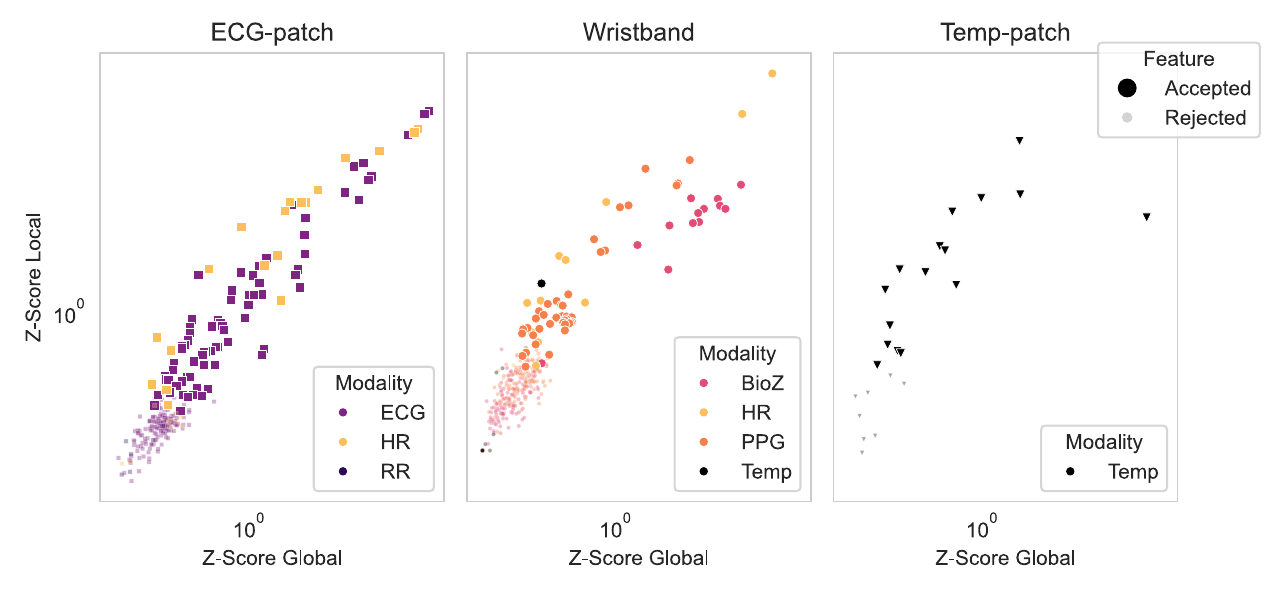}
    \caption{Device‐wise feature importances (z‐score) from BorutaSHAP. Each point is one feature. Modalities are illustrated in different colors. Accepted features are shown with big circles while rejected features are shown with smaller circles.}
    \label{fig:botura_sensor}
\end{figure}

Figure~\ref{fig:botura_sensor} shows global (horizontal) vs.\ local (vertical) z-scores of TreeSHAP importances from BorutaSHAP, stratified by device. Global scores reflect feature relevance across all modalities, while local scores capture relevance within each device. Features in the upper-right quadrant are informative both generally and device-specifically. The multimodal wristband produced the highest number of high-scoring features; the ECG-patch contributed predominantly to ECG and HR features; the Temp-patch offered fewer but consistent Temp features.

\begin{figure}[tb]
    \centering
    \includegraphics[width=\linewidth]{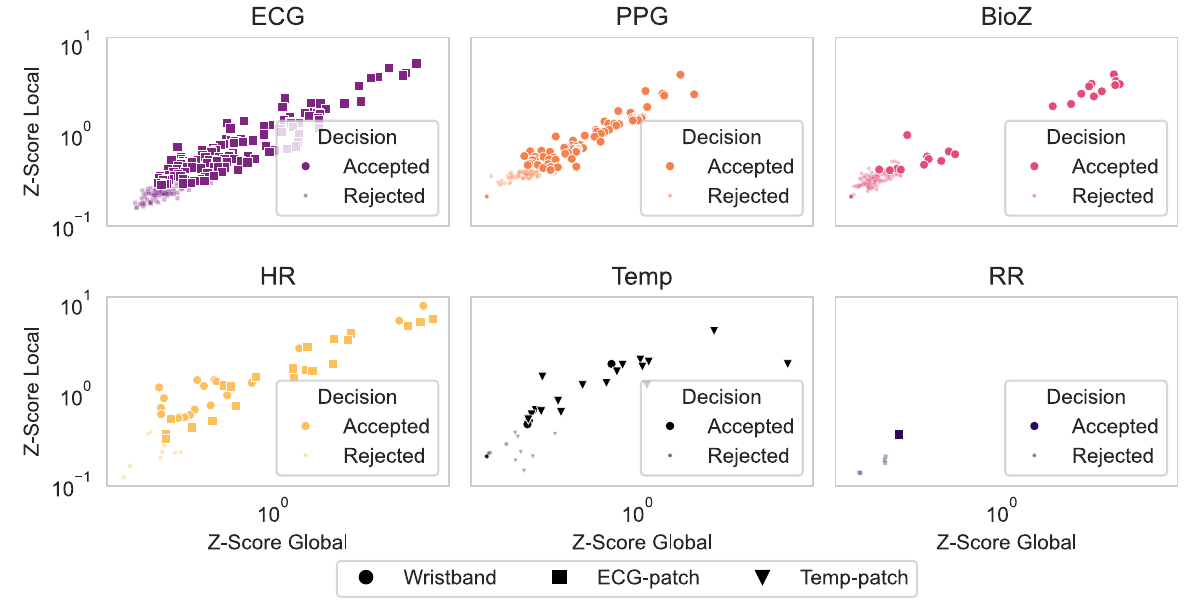}  %PREVIOUS: modality_wise_color
    \caption{Modality‐wise feature importances (z‐score) from BorutaSHAP.  Accepted features are shown as larger markers, while rejected features are shown as smaller markers. Maker forms indicates device.}
    \label{fig:boruta_modality}
\end{figure}

Figure~\ref{fig:boruta_modality} presents the same analysis grouped by modality. ECG and HR yielded the most accepted features with high global and local z-scores. BioZ and PPG also contributed several discriminative features, while Temp and RR had fewer overall. Among Temp features, those from the Temp-patch device showed higher importance; the Temp-patch provides both core and skin temperature, while the multimodal wristband measures only skin Temp. For HR, more accepted features were contributed by the ECG-patch than the wristband; the ECG-patch derives HR from ECG, whereas the wristband uses PPG.

\subsection{Ensemble Classifier Performance}
\begin{figure}[tb]
\centering
\includegraphics[width=\linewidth]{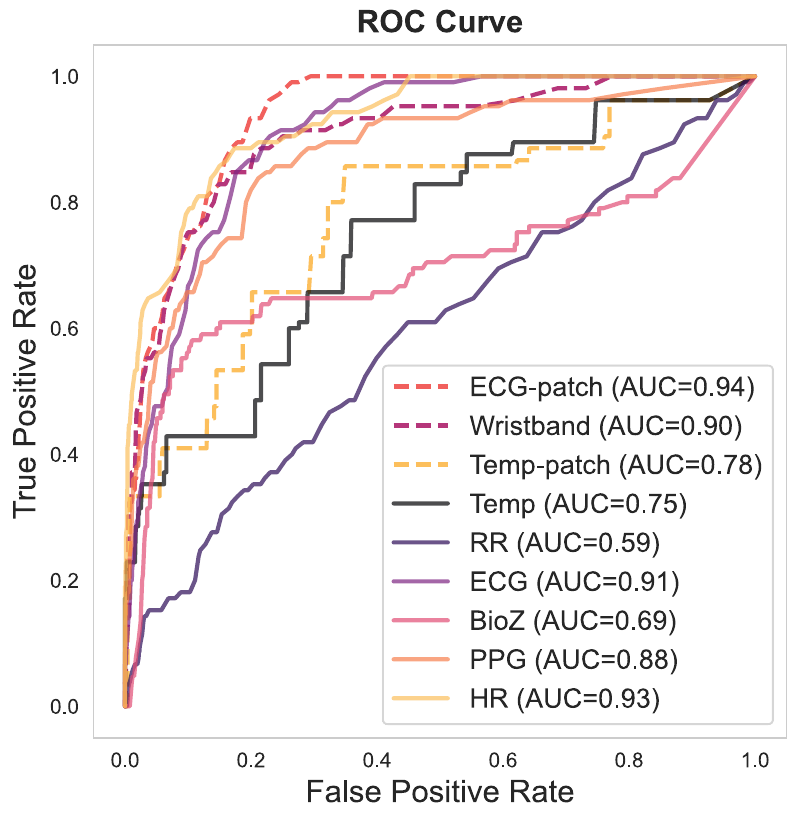}%
\caption{Receiver Operating Characteristic (ROC) curves of weak learners trained on individual signal modalities (Temp, RR, ECG, BioZ, PPG, HR) or device-specific feature sets (ECG-patch, multimodal wristband, Temp-patch). Curves correspond to a single run from the 10 repeated evaluations.}
\label{fig:roc}
\end{figure}

Figure~\ref{fig:roc} compares the ROC curves for each individual weak learner. Among the devices, the ECG-patch achieved the highest performance with an AUC of 0.94, followed by the multimodal wristband with an AUC of 0.90 and the Temp-patch with an AUC of 0.78. When analyzing performance by modality, HR yielded the highest AUC (0.93), closely followed by ECG (0.91) and PPG (0.88)%, highlighting their relevance in detecting AD-related cardiovascular changes
. Temp features showed a moderate AUC of 0.75, while BioZ and RR features had lower predictive value, with AUCs of 0.69 and 0.59, respectively.

\begin{figure}[tb]
    \centering
    \includegraphics[width=\linewidth]{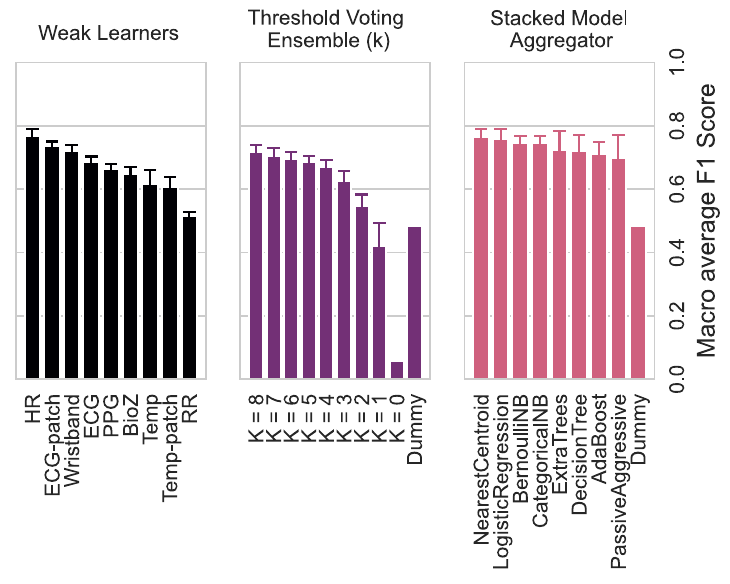} %allruns_R3_BAR_BAL.png
    \caption{Macro F1-scores of individual weak learners (left), $k$-threshold voting ensembles (middle), and stacked model aggregators (right). Learners include both modality- and device-specific models. Ensemble voting is shown across $k$ thresholds. The Dummy classifier serves as a baseline.}
    \label{fig:f1-score}
\end{figure}

Figure~\ref{fig:f1-score} presents macro F1-scores for all model components. The left panel shows individual weak learners, including modality-based (e.g., ECG, Temp) and device-based (e.g., ECG-patch, multimodal wristband) models, with HR and ECG-patch learners performing best. The middle panel shows $k$-threshold voting ensembles, with the highest performance at $k=8$ (0.72±0.02). The right panel compares the top stacked aggregators; Nearest Centroid achieved the top F1-score (0.77±0.03). A Dummy classifier yielded 0.48±0.00 as a baseline.

\begin{figure*}[]
    \centering
    \includegraphics[width=1.0\linewidth]{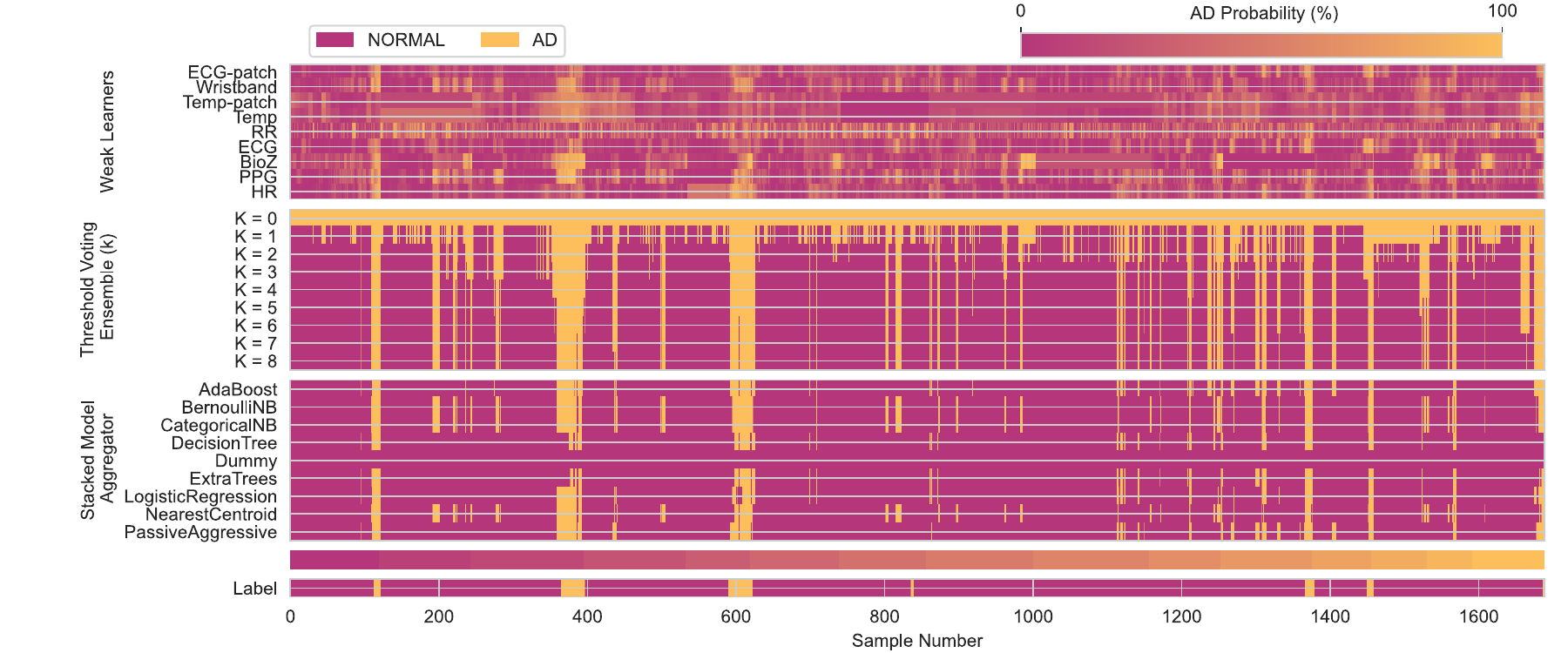}
    \caption{Temporal prediction overview across all samples of the dataset. Rows show AD prediction probabilities from: individual weak learners, $k$-Threshold Voting Ensembles ($k=0$ to $k=8$), stacked model aggregators, user IDs, and ground truth labels. Samples are ordered by subject and time. A Dummy classifier is included for baseline comparison.}
    \label{fig:chachiway}
\end{figure*}

Figure~\ref{fig:chachiway} shows the temporal progression of model predictions across the dataset. The top section displays probabilistic outputs of individual weak learners. The second shows $k$-Threshold Voting Ensembles ($k=0$ to $k=8$), where AD is predicted if at least $k$ learners agree. The third section shows outputs of selected stacked model aggregators. User IDs and ground truth labels are shown below. Samples are sorted chronologically per subject. This visualization supports inspection of consistency, model agreement, and alignment with AD episodes. It further enables qualitative assessment of detection performance in terms of correctly identifying AD, missing true events, over-predicting AD, and anticipating AD onset.

The results so far correspond to the configuration using all 9 weak learners, covering all modalities and devices. To assess robustness under sensor limitations, we evaluated all weak learner subsets. Table~\ref{tab:subset-comparisons} summarizes the five top-scoring combinations by macro F1-score. The highest F1 (0.78) was achieved by a reduced subset excluding Temp, suggesting that full inclusion is not strictly necessary. Several other subsets performed comparably (F1 = 0.75--0.77), indicating resilience to sensor loss. Notably, HR, ECG, and BioZ were common to the top-performing configurations, while PPG and Temp were more often excluded, suggesting they are less critical for stable AD detection.

\begin{table}[]
\centering
\caption{Classification metrics for the top-performing weak learner subsets. The full model (all 9 learners) achieves the best F1-score.
}
\begin{tabular}{p{0.55\linewidth} p{0.05\linewidth} p{0.05\linewidth}  p{0.05\linewidth} p{0.05\linewidth}}
\hline
Weak Learners & Acc & Prec & Rec & F1 \\
\hline
All                                 & 0.93 & 0.73 & 0.85 & 0.77 \\
ECG-patch, ECG, BioZ, PPG, HR, Temp-patch    & 0.92 & 0.71 & 0.82 & 0.78 \\
Wristband, ECG, BioZ, HR, Temp-patch         & 0.91 & 0.70 & 0.80 & 0.77 \\
Wristband, ECG, HR, Temp-patch, Temp         & 0.91 & 0.70 & 0.78 & 0.76 \\
ECG-patch, ECG, BioZ, HR, Temp-patch         & 0.91 & 0.70 & 0.79 & 0.76 \\
ECG, PPG, HR, Temp-patch                     & 0.89 & 0.68 & 0.76 & 0.75 \\
\hline
\end{tabular}
\label{tab:subset-comparisons}
\end{table}

\section{Discussion}
\label{sec:discussion}

\subsection{Principal Results}
%******** intro ******************************
\subsubsection{Data completeness} Figure~\ref{fig:data} highlights important considerations for real-world implementation. ECG-based HR signals from the ECG-patch were notably more reliable than PPG-based HR from the multimodal wristband, reinforcing the robustness of ECG-derived features for clinical use. Similarly, Temp recordings from the Temp-patch showed higher consistency than the wristband. Modalities like BioZ and RR demonstrated high availability across participants, while PPG and Temp from the wristband were more susceptible to dropout. These differences emphasize the importance of sensor selection in designing resilient multimodal systems, particularly when targeting deployment in uncontrolled or ambulatory environments.

\subsubsection{Windowing analysis} Figure~\ref{fig:windows_analysis} shows that longer window sizes (notably 60\,s) improved overall performance across weak learners, likely due to better spectral resolution and increased chances of capturing full AD episodes within a single window, thus reducing class imbalance. However, optimal window size varied by modality. Shorter windows favored Temp and RR, which change more slowly, while intermediate durations were best for BioZ, balancing noise reduction and responsiveness. Longer windows benefited ECG, PPG, and HR by better capturing rhythm-related and morphological features. These results support the adoption of modality-specific windowing strategies to optimize multi-modal AD detection.

% ******SHHAPLEY FEATURE DEPENDENCE discussion********
\subsubsection{Feature and Device Selection} As shown in Figure~\ref{fig:shap-analysis}, ECG- and HR-derived features dominated the top predictors (12 of 20), highlighting their importance in capturing cardiac dynamics associated with AD. This is consistent with studies linking HR changes—particularly reductions—to AD episodes~\cite{Yee_2022}, though variability has also been reported~\cite{Lindan_1980}. PPG features (4 of 20) were also informative, especially \textit{Tsystolic}, which reflects pulse wave propagation time and is inversely correlated with BP~\cite{Park_2021}. BioZ contributed two raw-signal-derived features (\textit{min}, \textit{1der-mean}), indicative of sympathetic arousal. A single temperature-based feature—mean second derivative over 3 minutes—ranked among the top, reflecting thermal acceleration patterns. No RR-based features appeared in the top 20. This diverges from earlier work~\cite{Suresh_2018}, which ranked GSR (analogous to BioZ) and Temp highest and HR lowest. The relatively lower contribution of Temp in our model may be partly due to missing data, according to Figure~\ref{fig:data}. Importantly, the aggregate of the remaining 612 features still accounted for a considerable share of the model’s predictive signal.

%% Which modalities?
By calculating BorutaSHAP z-scores globally and locally for either one modality or sensor (Figure~\ref{fig:boruta_modality}), we introduce a very robust statistical method for discarding weak features and choosing sensors or modalities with strong predictive power. This method is applicable and potentially useful for many other sensor selection tasks. The BorutaSHAP results (Figures~\ref{fig:botura_sensor}–\ref{fig:boruta_modality}) confirm ECG and HR as the most informative modalities, contributing the highest number of accepted features with strong global and local relevance. Among devices, the multimodal wristband and the ECG-patch yielded the most high-scoring features overall, while the Temp-patch contributed fewer but consistently relevant Temp features. Notably, the Temp-patch outperformed the wristband in Temp feature importance, likely due to its measurement of both core and skin temperature, whereas the wristband provides only skin Temp. Likewise, HR features derived from the ECG-patch were more informative than those from the wristband, suggesting that ECG-based HR estimation is more robust than PPG-derived HR in the context of AD detection.

%% Which exact features are shown to be relevant for AD (ECG, PPG)?
Using BorutaSHAP z-scores, we identified ECG and PPG feature clusters beyond simple HR features that were particularly descriptive. For ECG, notably well-performing are Hermite Basis Function Expansion (HBF) features, individual uniform Local Binary Pattern (uLBP) features, Higher Order Statistics (HOS), Heart Rate Variation and R-peak interval derived features. For PPG, especially systolic and diastolic peak amplitude durations calculated with the package BIOBSS, segment-based frequency and statistical features, and spectral and wavelet features show generalization across users.

% ******BORUTSAHP FEATURE DEPENDENCE discussion********
%******* AUC Scores********
The ROC analysis (Figure~\ref{fig:roc}) highlights the ECG-patch (AUC = 0.94) and the wristband (0.90) as best-performing devices, while the Temp-patch (0.78) lagged, likely due to its reliance on a single, less-informative modality (Temp), as also reflected in the SHAP results. Among modalities, HR achieved the highest AUC (0.93), followed by ECG (0.91) and PPG (0.88); RR showed the weakest performance (0.59), consistent with the BorutaSHAP findings (Figure~\ref{fig:boruta_modality}). According to established guidelines~\cite{Hosmer_2013,Mandrekar_2010}, these results reflect outstanding discrimination (AUC~$> 0.9$) for HR, and excellent performance (0.8--0.9) for ECG and PPG—supporting their clinical relevance for AD detection.

%******* F1-SCORES Scores********
The top meta-model using all weak learners was the Nearest Centroid Classifier (F1 = 0.77±0.03), substantially outperforming the Dummy baseline (F1 = 0.48±0.00) (Figure~\ref{fig:f1-score}). Yet, the best overall performance (F1 = 0.78) was achieved using a reduced set of modalities (ECG-patch, ECG, BioZ, PPG, HR, Temp-patch), suggesting that additional signals may introduce redundancy or noise (Table~\ref{tab:subset-comparisons}). The consistent exclusion of RR and wristband Temp reinforces the higher reliability of Temp-based features derived from the Temp-patch.

% Temporal evaluation
The temporal prediction map (Figure~\ref{fig:chachiway}) provides insight into model behavior during real-time AD events. Weak learners like HR and ECG aligned well with true AD episodes, supporting their reliability. In contrast, RR and Temp were more prone to false positives, with Temp often rising before AD onset. 
While this may lower precision, it suggests value for early warning and preemptive intervention. Missed detections also occurred, often for low-intensity episodes with subtle physiological changes. These findings highlight the complementary roles of modalities—early-reacting (e.g., Temp) and high-specificity (e.g., HR)—that could be leveraged in ensemble design to enhance both sensitivity and timing.

\subsection{Limitations}
A first limitation was defining a reliable BP baseline for AD detection in SCI is difficult due to high physiological variability~\cite{Deutges_2024}. Reported mean arterial pressure (MAP) variability while sitting is 17±4 mmHg (20\% MAP) in tetraplegia and 13±2 mmHg (12\%) in paraplegia; in recumbency, 13±3 mmHg (20\%) and 8±2 mmHg (8\%), respectively~\cite{Frisbie_2007}. We adopted the baseline definition from~\cite{Hubli_2014}, averaging the first three resting BP values before UDS. While practical, this may not capture circadian or lesion-specific variability, affecting labeling accuracy. 
Further, continuous reference labels were interpolated from sparse BP samples ($\geq$ min apart). 
Though enabling alignment with wearable data, this may misrepresent AD onset or duration. These limitations underscore the need for denser BP sampling and personalized baselines for improved detection accuracy.

Another key limitation is the small sample size. Although 27 participants were enrolled, only 17 were retained for analysis due to data quality issues, and just 7 exhibited AD during UDS. This limits the generalizability. These challenge highlights the need for coordinated efforts to build larger, high-quality, and publicly available datasets to advance AD research and clinical care in SCI.

While this study demonstrates the feasibility of AD detection using wearable sensors, the results  generalization to real-life ambulatory settings due to increased motion artifacts and uncontrolled conditions are yet to be tested. 
Nevertheless, the findings establish a critical proof-of-concept for wearable-based AD monitoring, motivating future work in free-living environments with enhanced artifact mitigation and signal robustness.

\subsection{Comparison With Prior Work}
Compared to prior studies, our work contributes with the first comprehensive analysis of non-invasive, and generalizable AD detection system with a focus on methodological rigor to understand physiological response that will lead to a predictive AD detection systems able to identify AD episodes before BP rises.

In contrast to \cite{sureshAutomatedDetectionSymptomatic2020,sureshAutomaticDetectionCharacterization2022}, which relied on user self-reports and lacked cross-validation, our model employs stratified cross-validation and a larger dataset to minimize overfitting and improve generalizability. Additionally, we introduce stricter ground-truth labeling criteria based on combined physiological patterns rather than subjective inputs.

While \cite{Pancholi_2024} demonstrated promising accuracy using deep learning on SKNA data in rat models, the invasive data acquisition and animal-based validation limit real-world deployment in human subjects. In contrast, our approach is fully non-invasive and evaluated directly on human participants.

Sagastibeltza et al. \cite{sagastibeltzaPreliminaryStudyDetection2022} used a highly controlled bladder-filling protocol to simulate AD but tested on only five individuals without wearable technologies or real-world conditions. Our model, trained on data collected during naturalistic daily activities, better captures the variability and unpredictability of community-dwelling individuals with SCI.

Finally, although \cite{sureshFeatureSelectionTechniques2022} explores important algorithmic improvements, it stops short of integrating these insights into a fully-deployed system with validation beyond training data. Our work operationalizes such optimizations into a scalable pipeline.

Overall, our contributions center on improving generalizability through stratified cross-validation and increasing population size, enhancing ecological validity via data collection from human SCI participants during real-world clinical procedures, and focusing on scalable, non-invasive monitoring solutions. Crucially, our study departs from earlier work by defining AD episodes using objective, continuous BP measurements rather than subjective self-reports or artificially induced events. This approach enables more accurate event labeling, minimizes bias, and strengthens the physiological validity of the model’s predictions.

\section{Conclusions}
This study successfully developed and validated a robust multimodal wearable-sensor system for automated, non-invasive AD detection in individuals with SCI, addressing a critical need for personalized cardiovascular monitoring. Our findings underscore the significant role of HR and ECG in accurately identifying AD episodes, providing reliable physiological markers for this life-threatening condition. The ensemble classification framework demonstrated resilience to sensor limitations, with key modalities maintaining strong predictive performance, which is vital for real-world application. This work represents a crucial step toward empowering individuals with SCI through continuous, personalized health surveillance, enabling earlier detection and proactive management of AD events. Future efforts will focus on optimizing these systems for everyday ambulatory use, further enhancing personalized health-care and preventive interventions.

%\appendix
%\section{Example Appendix Section}
%\label{app1}

%\appendix

% \section*{CRediT authorship contribution statement}

% %https://www.elsevier.com/researcher/author/policies-and-guidelines/credit-author-statement
% \textbf{Bertram Fuchs:} Methodology, Software, Investigation, Writing – review and editing. 
% \textbf{Mehdi Ejtehadi:} Data curation, Software, Investigation, Visualization, Writing – original draft.
% \textbf{Ana Cisnal:} Software, Validation, Visualization, Writing – original draft.
% \textbf{Jürgen Pannek:} Conceptualization, Resources, Writing – review and editing, Funding acquisition.
% \textbf{Anke Scheel-Sailer:} Conceptualization, Resources, Writing – review and editing.
% \textbf{Robert Riener:} Project administration, Supervision, Writing – review and editing, Funding acquisition.
% \textbf{Inge Eriks-Hoogland:} Conceptualization, Resources, Writing – review and editing.
% \textbf{Diego Paez-Granados:} Conceptualization, Methodology, Investigation, Project administration, Supervision, Writing – review and editing, Funding acquisition.

\section*{Declaration of Competing Interest}
The authors declare that they have no known competing financial interests or personal relationships that could have appeared to influence the work reported in this paper.

\section*{Data Availability Statement}
The datasets generated and analyzed during the current study are not publicly available due to participant privacy and institutional ethical restrictions. Data access may be considered upon reasonable request to the corresponding author and with appropriate ethical approvals.

\section*{Ethics Statement}
All procedures involving human participants were conducted in compliance with relevant Swiss laws, institutional guidelines, and the Declaration of Helsinki. The study protocol was reviewed and approved by the Ethics Committee of Northwestern and Central Switzerland (EKNZ), under the following reference numbers: 
\begin{itemize}
    \item EKNZ-2023-00400 (approved on 25.04.2023)
    \item EKNZ-2023-01050 (approved on 31.08.2023)
    \item EKNZ-2024-01254 (approved on 15.08.2024)
\end{itemize}

All participants provided written informed consent prior to participation, in accordance with Swiss Human Research Ordinance (HRO Art. 7, Risk Category A). Participant privacy rights were strictly observed, and all data were anonymized prior to analysis.

\section*{Declaration of generative AI and AI-assisted technologies in the writing process}
During the preparation of this work the author(s) used \textbf{ChatGPT by OpenAI} in order to enhance the clarity and fluency of the English language. After using this tool, the author(s) reviewed and edited the content as needed and take(s) full responsibility for the content of the publication.

\section*{Acknowledgments}
This study was funded by the Schweizer Paraplegiker Stiftung and the ETH Zürich Foundation (2021-HS-348), within the Digital Transformation in Personalized Healthcare initiative for individuals with spinal cord injury.
We would like to thank Mrs. S. Amrein, and the staff at SPZ Neurourology department for their work in data collection.


%%       \citet{<label>} ==> Lamport [21]
%%       \citep{<label>} ==> [21]
%%
%Example citation, See \citet{lamport94}.

\bibliographystyle{elsarticle-num-names} 
\bibliography{references}



%\begin{thebibliography}{00}


%\bibitem[Lamport(1994)]{lamport94}
%  \textit{\LaTeX: a document preparation system},
%  Addison Wesley, Massachusetts,

%\end{thebibliography}
\end{document}